# Effect of Topological Non-hexagonal Rings and Stone Wale Defects on the Vibrational Response of Single and Multi-Layer Ion Irradiated Graphene

Ashis K. Manna,[1,2] Simeon J. Gilbert,[3] Shalik R. Joshi,[1,2] Takashi Komesu,[3] Shikha Varma[1,2*]

[1] Institute of Physics, Sachivalaya Marg, Bhubaneswar Orissa 751005, India

[2] Homi Bhabha National Institute, Training School Complex, Anushakti Nagar, Mumbai 400085, India

[3]Department of Physics and Astronomy, Theodore Jorgensen Hall, 855 North 16th Street, University of Nebraska-Lincoln, Lincoln, NE 68588-0299, U.S.A.

*Corresponding Author: shikha@iopb.res.in

**Abstract**

Present study explores the observation of topological non-hexagonal rings (NHR) and Stone Wale (SW) defects by Raman experiments in both single (SLG) and multi-layer graphene (MLG) after they are irradiated with 100–300 eV Ar ions. Although predicted by theoretical studies, here it is experimentally shown for the first time that graphene SW/NHR defects have a signature in Raman. Broad bandwidth of the pertinent Raman features suggests the presence of more than one SW/NHR defect mode, in agreement with the DFT studies. Variations in the SW/NHR related Raman mode intensities demonstrate the annihilation of these topological defects at higher energies. Behavior of Raman allowed G and 2D excitations, as well as the disorder-activated D, D' and G* lines, has also been investigated in SLG and MLG. These indicate an evolution of defects in graphene with ion irradiation, as well as presence of a transition state beyond which the Raman modes are dominated by a rise in $sp^3$ content. Correlation of these aspects with the SW/NHR Raman provide significant insight into ion induced evolution of graphene. The direct observation of SW/NHR defects by Raman spectroscopy could be important in promoting exploration of rich topological aspects of Graphene in various fields.

## 1. Introduction

The functional properties of graphene highlight many unique aspects of the two-dimensional systems [1, 2] and have led to many frontier applications in the field of devices and sensors [1]. Development of the simplest electronic devices demands, however, knowledge of the defects [3]. Systematic inclusion of the defects in graphene may also induce many unique structural, vibrational, electronic, and optical properties [4-9]. On the theoretical front also, the nature of defects in graphene and carbon nanotubes show many rich characteristics [10-14] that demand more experimental understanding. The controlled introduction of defects, with electron or ions, provide technologically important platforms for many materials [15-18] that illustrate many important aspects and challenges in graphene [4-9, 19-21]. Formation of graphene quantum dots and the opening of electronic energy bandgap, with the introduction of disorder, present many fundamental and exciting possibilities for tailoring optical devices [1, 4]. The most common defects that are created in graphene, during knock-on events, are the point vacancies which are caused by the breaking of the three C-C bonds and involve about 8 eV energy [11, 12]. Mending of the vacancies, as suggested by some DFT studies, can also take place [10]. Formation of di-vacancies in graphene involve about 1eV more energy than the single vacancy [10].

Raman spectroscopy is a powerful technique to probe graphene since the π states associated with $sp^2$ bonds display significant resonance response to visible range photons [1, 22-25]. For the pristine graphene, Raman spectrum is dominated by the G band, of frequency 1582cm$^{-1}$, which originates due to the in-plane stretching of $sp^2$ atoms in the carbon-chains and rings. Introduction of point defects in graphene will scatter phonons, influencing the frequencies and asymmetries of the Raman allowed excitations. The disorder- activated Raman lines also appear in response to the impact. The D (defect) band, at the frequency of 1350 cm$^{-1}$, originates due to the breathing vibrations of the six membered C- rings in the presence of disorder [1, 23]. A second order process creates 2D mode with the conservation of momentum of two photons having opposite wave vectors. This mode, seen at the frequency of 2600 cm$^{-1}$, does not require any defect for activation [1, 23]. The ratio of Raman intensities of 2D and G bands are widely applied to distinguish single layer graphene (SLG) and multi- layer graphene (MLG). Another disorder related D' band [1, 23-25] originates with one phonon double resonance intra valley Raman process between an LO phonon, in the vicinity of Γ point, and a defect in the D' band. Introduction of energetic ions in graphene show many distinct disorder induced structural variations, but these usually involve increase of integral numbers of ions per area [5-9, 21], while the kinetic effects and the influence of energy of the bombarding species have received less attention [20]. Using molecular dynamic (MD) simulations, *Abdol et al.* have demonstrated the modulation in the interlayer linking between multilayer graphene sheets, with ion energy [21].

Topological defects in graphene, like non hexagonal rings (NHR) of pentagons and heptagons, or their combinations, like Stone Wale (SW) defects with pair of pentagons and heptagons, are of great interest and have been intensely investigated by theoretical methods [1,11,12,14]. SW defects form due to the rotation of a C=C bond, which transforms the four adjacent hexagons into two pentagons and two heptagons [1, 10-14]. As these defects introduce only local disturbance, with no long-ranged effects and their numbers are small, they do not produce significant visible deformation and so their identification in TEM is also difficult [1, 13]. The NHR defects form via single bond rotation and can get easily incorporated in the graphene lattice [10]. Methods like ion irradiation, chemical treatment and high temperature quenching are likely to form SW/NHR defects [1, 14]. These defects require a low free energy of formation (3.5 eV) compared to the mono-vacancy or di-vacancy [1, 10-13]. Many theoretical studies, using first principle calculations and DFT, have proposed characteristic Raman peaks for SW/NHR defects [11, 13, 14]. Although such defects are prevalent in ion irradiated graphene, so far these defects have not been observed in Raman experiments. One of the reasons is their weak Raman intensity, additionally, several of the signatures overlap with other prominent Raman bands, making their delineation hard. *Shirodkar et al.* have used the first principles analysis to show that the frequencies of the G and D bands get effected by the concentration of the SW defects [12]. The direct observation of SW/NHR defects by Raman spectroscopy could be extremely important in fundamental studies of topological defects as well as for exploring their vibrational and dynamical aspects in graphene.

In the present study we utilize Raman spectroscopy to explore the evolution of vibrational properties of Single layer Graphene (SLG) and Multi-layer Graphene (MLG) after the introduction of systematic disorder, through irradiation with the low energy $Ar^+$ ions of 100-300 eV at a constant fluence of 2.2 x $10^{14}$ ions/cm$^2$. Significantly, a weak Raman band near 1800-2200 cm$^{-1}$ is observed in ion irradiated SLG and MLG. This is intriguing since theoretical studies, based on first principle calculations and DFT, have predicted a mode for topological NHR/SW defects in this range. However, this mode so far has not been observed or investigated in Raman experiments. Both in SLG and MLG, this Raman band is very broad indicating the presence of multiple SW/NHR related modes, as also suggested by DFT studies. The intensity behavior reflects the annihilation of these defects at high energy. The evolution of G and 2D Raman bands, as well as defect activated D, D' and G* has also been investigated with ion energy and reflects the presence of a transition stage ($T_s$), beyond which $sp^3$ content rises and becomes a dominant component. Significance and correlation of the SW/NHR defects with ion energy presents some fascinating aspects. The direct observation and evolution of these topological defects in Raman experiments can provide very important knowledge for understanding topological defects in graphene.

## 2. Experimental Section
### 2.1 Sample preparation:

Graphene samples were utilized as-received from Graphenea (Spain). Graphene films were implanted with $Ar^+$ ions at normal incidence at room temperature under the vacuum of 5 x 10$^{-6}$ Torr. Ion implantation was carried out at several ion energies of 100, 150, 200, 250 and 300 eV. During implantation, the fluence and the flux were kept constant at 2.2 x $10^{14}$ ions/cm$^2$ and 3.7 x $10^{12}$ ions/cm$^2$·sec, respectively. In order to obtain a uniform implantation, the ion beam was scanned over the entire sample. The lowest ion energy used here is 100 eV. This was selected as it corresponds to the energy required for overcoming the C atom displacement barrier, without introducing any cascading effect [26].

### 2.2 Characterization:
The Raman measurements have been undertaken in the backscattering geometry on a T64000 triple monochromator (Horiba Jobin Yvon) having a resolution of 0.5 cm$^{-1}$. The system is equipped with a liquid nitrogen cooled CCD detector. All the spectra presented here have been acquired with a 514 nm (2.41 eV) laser with a low power of less than 1 mW. Raman spectra were fitted with Lorentzian profiles.

XPS measurements were undertaken with an Al K$\alpha$ source (1486.6 eV) in an ultra-high vacuum (UHV) system with a VG100AX hemispherical analyzer. Spectra were acquired at a normal photoelectron emission angle, with respect to the sample surface. The base pressure in the UHV system was maintained at better than 1 x 10$^{-9}$ Torr. Surface morphology was investigated with a Nanoscope V (Bruker) Atomic Force Microscope (AFM).

## 3. Results and Discussion:

The Raman spectra of pristine as well as ion irradiated graphene are shown for MLG in Fig. 1. For the pristine MLG, G and 2D Raman modes are seen at 1583.5 and 2680.5 cm$^{-1}$, respectively

[1, 23], while a tiny D mode is seen at 1345 cm$^{-1}$ which is attributed to the presence of grain boundary and edges in graphene [1, 20, 22].

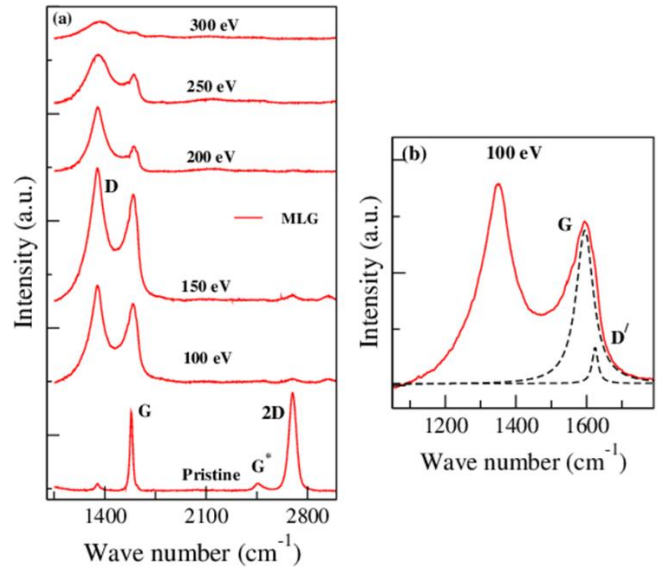

**Figure 1:** *(a) Raman spectra after irradiation of graphene at various ion energies for MLG. The ion irradiation fluence was kept constant at 2.2 x 10$^{14}$ ions/cm$^2$. (b) Raman spectrum from MLG, after irradiation at 100 eV displaying deconvolution of the G and D' modes.*

Ion irradiation primarily introduces vacancy type defects which induce several changes in the lattice (fig. 1). Most prominent among these are the decay of the 2D mode, and modifications in the G and D modes. In addition, some other important defect related modes are also observed that are weaker in intensity. After 100 eV irradiation, a minor D' mode appears as a shoulder to the G band (fig. 1b). Inclusion of defects contribute to the phonon states with nonzero q momentum, leading to the modes like D and D' which are not present in crystalline, defect free graphene [1, 4, 5, 9, 23]. Near the 2D band, at a slightly lower frequency, a small G$^*$ band is present that originates via an inter valley double resonance (DR) process that involves two phonons near the K-point: an in-plane longitudinal acoustic (iLA) and in-plain transverse optical (iTO) phonon. Ion irradiation of SLG, which shows fabrication of quantum dots, also introduces many phonon defect states [4].

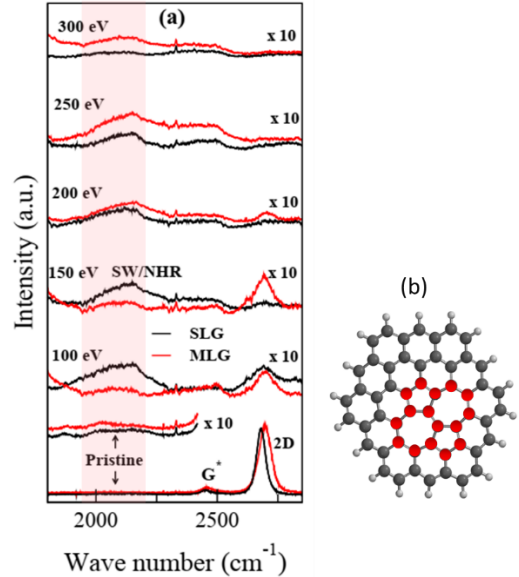

**Figure 2:** *(a) Raman spectra after irradiation of graphene at various ion energies for SLG (black lines) and MLG (red lines). The ion irradiation fluence was kept constant at 2.2 x 10$^{14}$ ions/cm$^2$. The shaded region (1800-2200 cm$^{-1}$) shows SW/NHR defect related Raman in irradiated SLG and MLG. Pristine is also shown for comparison and does not show SW/NHR defect (b) schematic of the SW/NHR defect (in red).*

*Popov et al.* have presented a detailed DFT study, based upon non-orthogonal tight binding model, predicting several Raman lines (at 1346, 1524, 1595, 1740, 1927, and 2142 cm$^{-1}$) as signatures of SW modes [11]. The structure of a SW defect is flat and these phonons originate due to the in-plane displacements. Among the six Raman excitations, the lower four originate from the displacements in heptagons, while the highest two by the pentagons. However, the Raman line at 2142 cm$^{-1}$ was proposed as the most likely candidate, being the most intense and well separated mode, away from other prominent graphene excitations [11]. For topological non hexagonal rings (NHR), like pentagon and heptagon defects, and also SW defects, characteristic vibrational modes have also been derived using an empirical bond polarizability model [13]. Using these first principle calculations it was proposed that the pentagon defects will produce characteristic peaks above the G peak, in frequency range 1820-2000 cm$^{-1}$, while the heptagon defects will introduce Raman lines below the G mode. Furthermore, a direct correlation of the number of pentagon defects with the intensity of the Raman peak was also suggested [13].

Remarkably, for the irradiated MLG, a very faint and broad Raman band is seen (in fig. 1) near 2000 cm$^{-1}$. Figure 2 shows a expanded section (from fig. 1) around this region. Also included here are the Raman spectra from ion irradiated SLG. Both SLG as well as MLG display a clear band in the frequency range 1800-2200 cm$^{-1}$, after ion irradiation. This Raman excitation is not present in the pristine graphene and is only observed after irradiation, attributing it to the incorporation of defects in the lattice. Based on the first principle studies [11, 12, 13], this band can be attributed to vibrational modes due to SW/NHR defects. Significantly this band has never been experimentally probed or investigated, as far as we are aware, probably because its intensity is usually very weak, when compared to G, D or 2D Raman modes. This fact is also noted in first principle studies [11]. High degradation of 2D mode, after irradiation in the present study, is helpful in exposing the SW/NHR feature in SLG as well as MLG.

In the following, the evolution of the dominant allowed Raman modes (G and 2D) is briefly discussed before presenting details of the disorder activated D, D' and G* modes. We pay special attention to the evolution of the SW/NHR modes. The Raman intensity, position and the full width at half maxima (FWHM) have been derived from the Lorentzian fitting of the respective modes. As the G and D' modes nearly overlap, they have been de-convoluted (fig. 1b) for obtaining individual contributions [1, 24].

*3.1 Evolution of Raman G and 2D mode with ion energy*

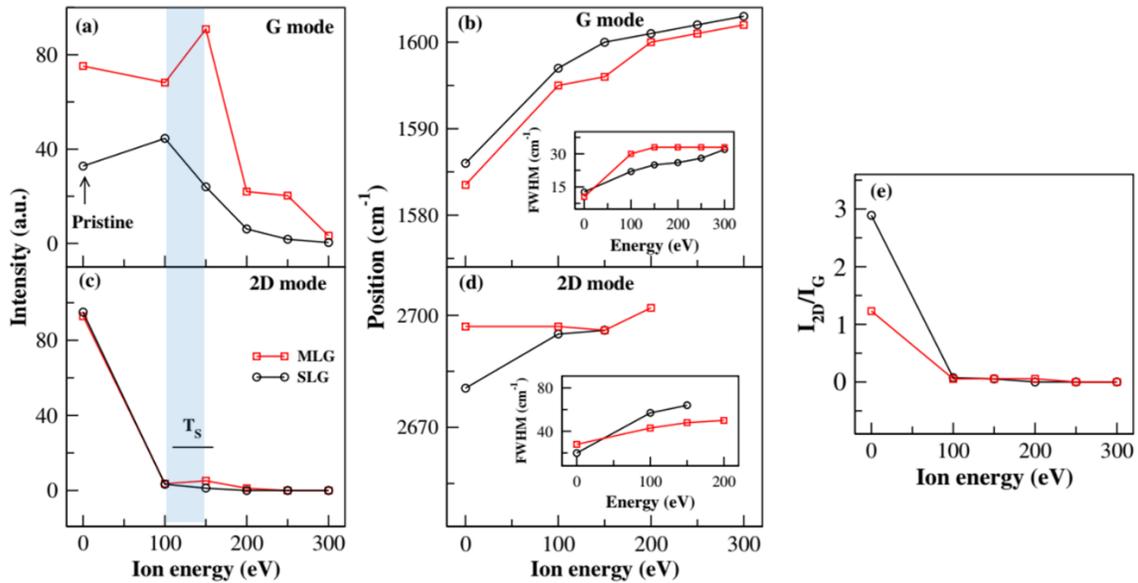

**Figure 3:** *G mode (a) Intensity and (b) Position (inset shows FWHM). 2D mode (c) Intensity and (d) Position (inset shows FWHM). (e) Ratio of $I_{2D}/I_G$. Data is shown for SLG (black lines) and MLG (red lines) after implantation at various ion energies, fluence was kept constant at 2.2 x $10^{14}$ ions/cm$^2$.*

The evolution of the G and 2D modes, after ion impact display many important characteristics (fig. 3). Several studies have investigated these modes with increasing ion fluence [1]. Here, the role of kinetic effects with increasing ion energy are emphasized, while the fluence was kept constant (2.2 x $10^{14}$ ions/cm$^2$). The evolution of the G mode intensity, for SLG and MLG with ion energy, (fig. 3) displays two main trends: a mild increase initially which subsequently degrades with ion energy. Some ion energy induced effects in the intensities of G and D mode have been earlier investigated for SLG while exploring graphene quantum dots [4]. The higher overall G intensity of MLG compared to SLG (in fig. 3) reflects its higher damage resistant nature. This is due to the fact that the Van der Waals interactions among multiple layers provide a greater rigidity against degradation [6]. Although the absolute numbers in intensity are arbitrary, the degradation initiates in the range of 100-150 eV, for both SLG and MLG. This is also compounded by an intense hardening of the mode due to the inclusion of compressive forces and structural distortions. Beyond 100eV in SLG and 150 eV in MLG, the mode consistently declines and only very mild hardening (in fig. 3b) is seen. This suggests the presence of a transition stage between 100 and 150 eV (shown as blue shaded region and labelled as $T_s$). Role of penetration depth is distinctly depicted by the FWHM, which is influenced differently by the number of layers: changing nearly linearly with energy for single layer but not showing much variation above $T_s$ in multi-layers. Quantum confinement effects due to the creation of graphene quantum dots can also induce widening [4]. At the highest (300 eV) impact energy, all the effects appear nearly independent of the number of layers. DFT studies indicate G-mode softening due to SW/NHR defects [12]. Presence of such defects here, as discussed in following sections, may also contribute to the decreased hardening beyond 150 eV (>$T_s$).

The high intensity of the 2D band (fig. 3c) as well as high $I_{2D}/I_G$ ratios of 2.8 and 1.2 for pristine SLG and MLG, respectively, confirms the existence of good quality single and multi-layer graphene [1, 23]. Ion impact causes a swift degradation of the 2D mode, even with just 100 eV irradiation. The mode exhibits hardening and broadening which increase with energy (Fig. 3d). The mode nearly vanishes at 200 eV and 250 eV, for single layer and multi-layers, respectively. The FWHM of the 2D mode in MLG is broader in pristine (30 cm$^{-1}$) but is less effected with energy.

*3.2 Evolution of Defect related Raman D and D' modes with ion energy:*

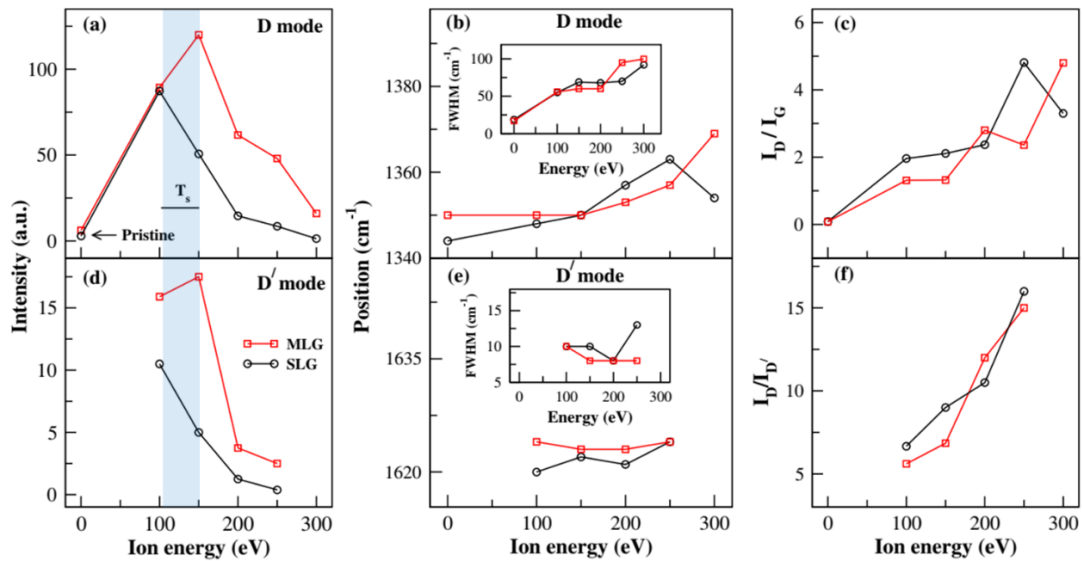

**Figure 4:** *D mode (a) Intensity, (b) Position (inset FWHM) and (c) Ratio $I_D/I_G$. D' mode (d) Intensity, (e) Position (inset FWHM) and (f) Ratio $I_D/I_{D'}$. Data is shown for SLG (black lines) and MLG (red lines) after implantation at various ion energies, fluence was kept constant at 2.2 x $10^{14}$ ions/cm$^2$.*

Both SLG and MLG display an initial increase in the disorder D mode intensity when impacted with ions. The intensity rises up to 100 and 150 eV, respectively, in SLG and MLG while it declines subsequently (fig. 4). The initial increase reflects the inclusion of vacancies and disorder in the lattice [1, 23]. At energies higher than 150 eV ($>T_s$), some deconstruction of the six-membered carbon–rings into lower ordered rings, or their clusters, can contribute to the degradation of this mode [5, 7]. The influence of ion energy is also indicated by the increased FWHM broadening which, however, is not affected by the number of layers. Observed at the positions 1345 and 1350 cm$^{-1}$ in pristine SLG and MLG, respectively, this mode does not shift much initially, but displays hardening at higher energies above $T_s$, indicating severe lattice distortions in the C-rings. These results are in contrast to the fluence dependent studies performed with 90 eV ions [5], where no change in the peak frequency was observed, even after amorphization at the fluence of 1 x 10$^{15}$ ions/cm$^2$. This demonstrates the significant role of the ion energy and the associated kinetic effects in understanding the lattice evolution with disorder, and emphasizes that fluence variation alone, as is the case in most of the earlier studies, does not capture the full scenario. The presence of grain boundary edges as well as SW/NHR defects can also influence the D mode. DFT results have shown a nearly two-fold increase in the D mode intensity near the carbon nanotube edges that are rich in such defects [10]. Hardening of the D mode, in the presence of SW defects, has also been reported [12].

The ratio $I_D/I_G$ displays a nearly similar rising nature for SLG and MLG suggesting an increase in ion induced disorder in the lattice with ion energy (fig. 4c). The ratio for MLG at 100 eV corresponds well with the earlier study where single energy (90 eV) for Ar ions at various fluences was used [6]. The ion energy induced effects on $I_D/I_G$ in SLG have been discussed earlier [4]. Here the comparison with MLG, however, indicates a usually higher $I_D/I_G$ for SLG (fig. 4c). Investigating SLG alone, *Ahlberg et al.* observed a nearly constant ratio (~ 1.1) after irradiation at 100 and 200 eV, the energy investigated in [20]. Not only this ratio is almost twice at these energies here (fig. 4c), it overall increases with ion energy exhibiting kinetic effects. The destruction of the sp$^2$ rings, during ion impact, contributes to the formation of damaged zones. Aggregation of such damaged areas decreases the crystallite size and leads to the increase in the $I_D/I_G$ ratio. During this process, sp$^3$ amorphous content also enhances and at some stage amorphization occurs. Decline in ratio at 300 eV for SLG reflects this stage of amorphization [4, 5, 7, 9], as also suggested by the softening of the D mode seen here (fig. 4a). AFM and XPS results, presented in the following sections reflect the aggregation of the damage as well as the formation of sp$^3$ defects after ion irradiation.

The D' band is also a disorder related mode and is absent in pristine. It is observed in ion irradiated SLG as well as MLG, being most intense at 100 and 150 eV, respectively. Its position and the width do not indicate much variation. It has been proposed that $I_D/I_{D'}$ is sensitive to the type of defects present in the lattice [24, 25] with ratios of 3.5, 7 and 13 suggesting edge defects, vacancies and sp$^3$ defects, respectively [25]. The ratio $I_D/I_{D'}$ is shown for SLG as well as MLG (fig. 4f) and vacancy type defects appear to be prevalent in the energy range of 100-150 eV, while sp$^3$ hybridized carbon networks are dominating in 200-250 eV regime. These observations, overall in concurrence with the $I_D/I_G$ ratio reflect a similar scenario for single and multilayer graphene. However, a very high $I_D/I_{D'}$ value (~15) but reasonably lower $I_D/I_G$ (~2) in MLG, at 250 eV energy, suggests that more studies in this regard are needed.
The positional shifts of Raman modes show some interesting character due to the kinetic effects. The G mode undergoes a distinct blue shift (Fig. 3b), within 100-150eV energy range, while the D mode gets only mildly affected (Fig. 4b). On the other hand, this situation gets reversed above 150 eV; with the G mode getting less influenced than the D mode. With C-C stretching being responsible for the origin of G mode whereas breathing of six-membered rings leading to the D mode activation, it appears that initially at lower energies, modifications primarily influence the stretching mode, whereas at higher irradiation energies (>$T_s$), C-ring breathing modes get significantly distorted. Formation of shorter and stiffer C-C bonds in SW defects can also produce softer G and harder D modes [11, 12]. Compared to the D mode, the D' mode does not show much positional variation.

*3.3 Evolution of SW/NHR and G* mode with ion energy:*

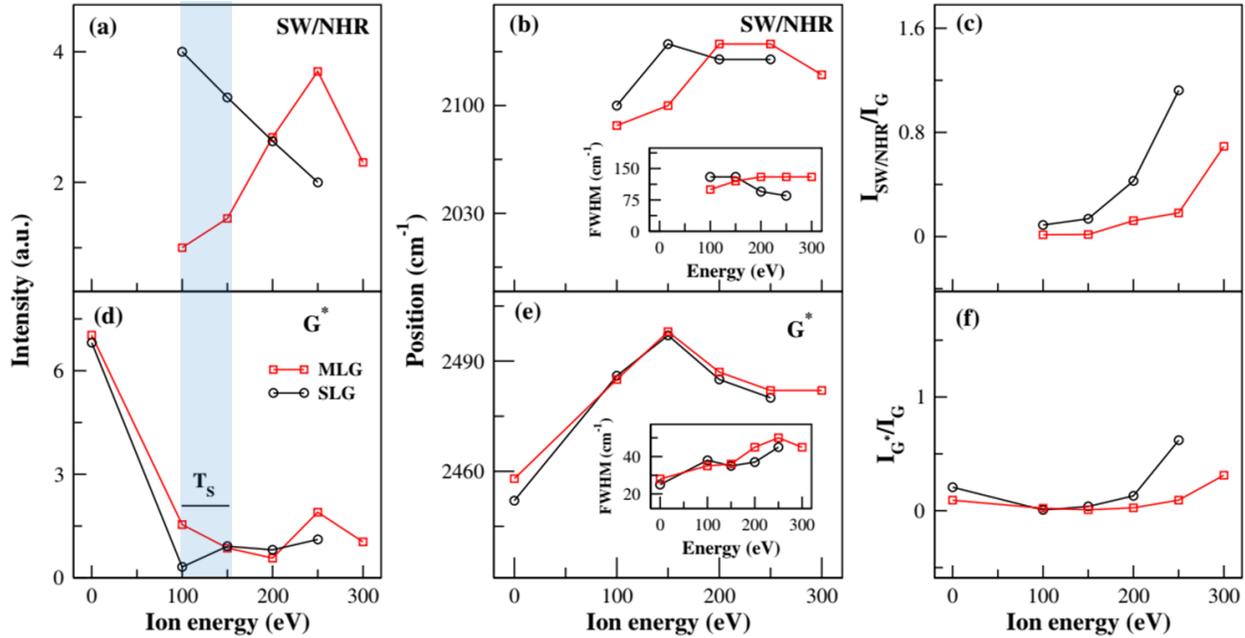

**Figure 5:** *SW/NHR mode (a) Intensity, (b) Position (inset FWHM) and (c) Ratio $I_{SW/NHR}/I_G$. G\* mode (d) Intensity, (e) Position (inset FWHM) and (f) Ratio $I_{G*}/I_G$. Data is shown for SLG and MLG after implantation at various ion energies, fluence was kept constant at $2.2 \times 10^{14}$ ions/cm$^2$.*

Figure 5 presents the evolution of SW/NHR band, seen in the frequency range 1800-2200 cm$^{-1}$ (in fig. 2) for ion irradiated SLG and MLG. According to the molecular dynamic simulations, formation of NHR defects takes place via the transformation of vacancy defects, during the saturation of dangling bonds [10]. Due to the linkage and Van-der Waals interactions between the layers [21], it will be more difficult to initiate bond rotations in MLG, necessary for forming SW/NHR defects. This can explain the much lower SW/NHR defect intensity at 100 eV in MLG, compared to SLG. The intensity behavior, with ion energy, is also very different for SLG and MLG. Observed first after 100 eV irradiation, the intensity systematically decays with increasing energy in SLG. On the other hand in MLG, the intensity rises up to 250 eV and declines only at the highest energy of 300 eV (fig. 5a). Kinetic Monte Carlo simulations have suggested the migration of embedded pentagonal rings within the graphene layer as well as on its edges [27]. Further, such diffusing non hexagonal rings can transform into six-membered rings, through isomerization at the graphene edges [27]. Enhanced SW/NHR defect migration, with increasing ion energy, and their subsequent annihilation, by getting transformed into hexagonal rings, may be responsible for the declining intensity in SLG (fig. 5a). In contrast, as the migration of SW/NHR defects will be more restrictive in MLG due to the interlayer interactions, the decay appears to be significant only at the highest energy. The overall positional variations of SW/NHR mode are nearly similar in SLG and MLG, with a hardening at low energies and a subsequent mild softening. The large bandwidth of this mode suggests presence of more than one SW/NHR mode within the band (shaded region of fig. 2), in support of earlier DFT studies [11, 13]. Higher resolution Raman measurements will be necessary to resolve these components. Although the width in MLG does not change much, with ion energy, it reduces significantly in SLG at higher ion energies. At the highest energy the SW/NHR mode is absent in SLG while it is very weak in MLG. The variation in ratio $I_{SW/NHR}/I_G$ with ion energy (fig. 5c) is small at low energies, both for SLG and MLG, but displays a mild rise above 200 eV. Domination of $sp^3$ defects will affect the presence of SW/NHR defects at this stage. Many more studies of this topological SW/NHR mode will be needed to elucidate the rich characteristics.

The intensity of G* mode rapidly declines with energy, showing a very different behavior than SW/NHR defect in SLG and MLG. The associated initial hardening but subsequent softening, as well as the broadening of G* are presented in Fig. 5. These also exhibit very different nature compared to the SW/NHR mode. The ratio $I_{G*}/I_G$ usually remains low.

*3.4 Electronic Structure and Morphology*

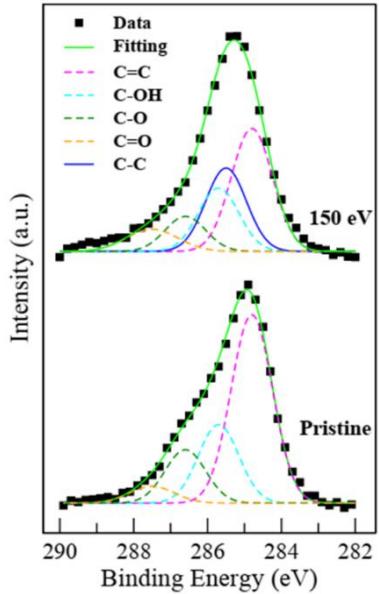

**Figure 6:** *C (1s) core level spectra for pristine and after irradiation with 150 eV ions. Ion fluence was kept constant at 2.2 x $10^{14}$ ions/cm$^2$.*

XPS measurements have been utilized to understand the modifications in the chemical species after ion irradiation. The C (1s) core level has been investigated (fig. 6) to study the presence of $sp^2$ and $sp^3$ bonded carbon species as well as the existence of any oxygen functional groups on the graphene surface. The pristine graphene exhibits an intense $sp^2$ feature corresponding to C=C

(284.8 eV) denoting the presence of good graphene layer. In addition, components related to C-OH, C-O and C=O are also observed at 285.7, 286.6 and 287.5 eV, respectively [28]. After irradiation, a new feature is noticed at 285.5 eV and is shown in Fig.6 for 150 eV ion energy. This feature is assigned to the C-C $sp^3$ defects, created due to the damage in $sp^2$ bonded carbon-rings [29].

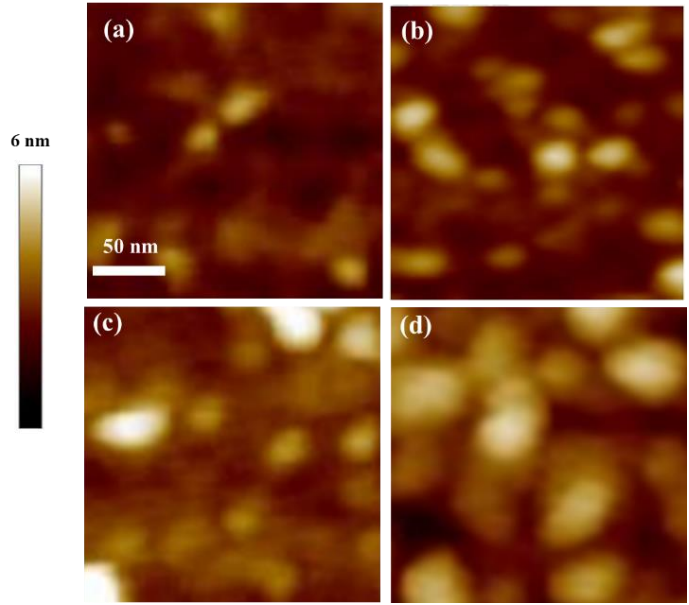

**Figure 7:** *High resolution 200 x 200 nm$^2$ AFM image from graphene after irradiation with (a) 100 eV (b) 150 eV (c) 200 eV and (d) 300 eV. Ion fluence was kept constant at 2.2 x 10$^{14}$ ions/cm$^2$.*

AFM images present the surface morphology of the ion irradiated graphene. After irradiation at 100 eV (fig. 7a), several weak protrusions attributed to the defects can be seen. Similar hillocks have been observed by *Lucchese* et al. in [5] where a single ion energy of 90 eV Ar$^+$ was used. The disorder can be attributed to the formation of $sp^3$ regions and distortions in the $sp^2$ rings [30]. After 150 eV irradiation, in addition to the increase in their density, the defects appear well defined and larger (fig. 7b). As some defects now overlap, the size of the undamaged graphene reduces. This results in the decay of the G and D Raman modes (fig. 3 and 4). While the dimensions of the $sp^2$ crystallite diminish, the $I_D/I_G$ ratio in Raman rises. Phenomenological studies on nano-crystalline graphite have also indicated this inverse relationship between $I_D/I_G$ and the crystallite dimensions [4, 5]. Drastic enhancement in the size of defects along with a significant coalescence of disorder occurs at 300 eV irradiation. This can induce high $sp^3$ content in the lattice, as suggested by fig. 4 (c and f).

**4. Conclusion:**

The present work illustrates the role of kinetic effects of ion irradiation energy (100-300 eV) on the vibrational response of the single layer and multilayer graphene. It is fascinating that the vibrational Raman response also exhibits the presence of topological non hexagonal ring/Stone Wale defects (NHR/SW) in the frequency range of 1800-2200 cm$^{-1}$. Such modes have been predicted by theoretical methods, however, as far as we are aware, these have never been probed by Raman experiments. The wide FWHM of NHR/SW defects demonstrates the presence of more than one frequency component, as proposed by earlier DFT studies. The decay in their intensity, at higher energies, reflects the transformation of these defects into six-membered rings. The evolution of G, 2D as well as disorder related D, D' and G* modes has also been investigated. These indicate the presence of a transition stage in 100-150 eV energy range. Several crucial modifications occur, beyond this transition stage, due to the increase in sp$^3$ content in the lattice. The evolution of these other Raman modes, G, 2D, D, D' and G*, in correlation with NHR/SW defects provide a rich picture of the role of ion energy. The direct observation of a NHR/SW mode by Raman experiment can be significant in fundamental fashion for the study of Topological defects in graphene as well as for understanding the physics and chemistry of the topological defects in general.

**Acknowledgements:**

S.V. and A.M. would like to acknowledge the financial help received from the Department of Science and Technology- Science and Engineering Research Board (DST-SERB), India (No. DST/EMR/2016/000728). This work was also partly funded by the Nebraska Public Power District through the Nebraska Center for Energy Sciences Research at the University of Nebraska-Lincoln, NCESR grant number 19-SE-2018. We sincerely thank Prof. Peter Dowben for very valuable suggestions and comments. We also thank Prof. Ajit M. Srivastava for useful discussions. Prof. T. Som is acknowledged for providing ion irradiation setup. Help of Mr. Santosh Choudhury with Raman set up is acknowledged.

**References**

1. Wu JB, Lin ML, Cong X, Liu HN, Tan PH. Raman spectroscopy of graphene-based materials and its applications in related devices. Chem Soc Rev 2018; 47(5):1822-1873.

2. Zhou M, Pasquale FL, Dowben PA, Boosalis A, Schubert M, Darakchieva V, Yakimova R, Kong L, Kelber JA. Direct graphene growth on $Co_3O_4$ (111) by molecular beam epitaxy. J Phys Condensed Matter 2012; 24(7): 072201.

3. Chen JH, Jang C, Xiao S, Ishigami M, Fuhrer MS. Intrinsic and extrinsic performance limits of graphene devices on $SiO_2$. Nat Nanotech 2008; 3: 206.

4. Manna A, Gilbert SJ, Joshi SR, Komesu T, Dowben PA, Varma S. Tuning photo-response and electronic behavior of graphene quantum dots synthesized via ion irradiation. Phys B: Condensed Matter 2021; 613: 412978.

5. Lucchese MM, Stavale F, Ferreira EHM, Vilani C, Moutinho MVO, Capaz RB, Achete CA, Jorio A. Quantifying ion-induced defects and Raman relaxation length in graphene. Carbon 2010; 48(5): 1592-1597.

6. Jorio A, Lucchese MM, Stavale F, Ferreira EHM, Moutinho MVO, Capaz RB, Achete CA. Raman study of ion-induced defects in N-layer graphene. J Phys Condensed Matter 2010; 22(33): 334204.

7. Zhou YB, Liao ZM, Wang YF, Duesberg GS, Xu J, Fu Q, Wu XS, Yu DP. Ion irradiation induced structural and electrical transition in graphene. J Chem Phys 2010; 133(23): 234703.

8. Ferrari AC, Robertson J, Interpretation of Raman spectra of disordered and amorphous carbon. Phys Rev B 2000; 61(20): 14095-14107.

9. Guo B, Liu Q, Chen E, Zhu H, Fang L, Gong JR. Controllable N-Doping of Graphene. Nano Letter 2010; 10(12): 4975-4980.

10. Collins PG, Narlikar AV, Fu YY, editor. Defects and Disorder in Carbon nanotubes, vol. 2. Oxford University Press; 2010. p. 73.

11. Popov VN, Henrard L, Lambin P. Resonant Raman spectra of graphene with point defects. Carbon 2009; 47(10): 2448-2455.

12. Shirodkar SN, Waghmare UV, Electronic and vibrational signatures of Stone-Wales defects in graphene: first-principles analysis. Phys Rev B 2012; 86(16): 165401.

13. Wu G, Dong J, Raman characteristic peaks induced by the topological defects of carbon nanotube intramolecular junctions. Phys Rev B 2006; 73(24): 245414.

14. Tailor PM, Wheatley RJ, Besley NA. An empirical force field for the simulation of the vibrational spectroscopy of carbon nanomaterials. Carbon 2017;113: 299-308.

15. Solanki V, Majumder S, Mishra I, Dash P, Singh C, Kanjilal D, Varma Shikha. Enhanced anomalous Photo-absorption from $TiO_2$ nanostructures. J Appl Phys 2014; 115: 124306.

16. Manna AK, Barman A, Joshi SR, Satpati B, Dash P, Chattaraj A, Srivastava SK, Sahoo PK, Kanjilal A, Kanjilal D and Varma Shikha. The effect of Ti+ ion implantation on the

Anatase- Rutile Phase Transformation and Resistive Switching properties of TiO2 thin films. Jour. Appl. Physics 2018; 124: 155303.

17. Joshi SR, Padmanabhan B, Chanda A, Malik VK, Mishra NC, Kanjilal D, Varma Shikha. Effects of cobalt implantation on structural and optical properties of rutile $TiO_2$ (110). Appl Phys A 2016; 122(7): 713.

18. Dey S, Roy C, Pradhan A, Varma Shikha. Raman scattering characterization of Si (100) implanted with mega electron volt Sb. J Appl Phys 2000; 87(3): 1110-1117.

19. Meyer JC, Kisielowski C, Erni R, Rossell MD, Crommie MF, Zettl A. Direct imaging of lattice atoms and topological defects in graphene membranes. Nano Letter 2008; 8(11): 3582-3586.

20. Ahlberg P, Johansson FOL, Zhang ZB, Jansson U, Zhang SL, Lindblad A, Nyberg T. Defect formation in graphene during low energy ion bombardment. APL Materials 2016; 4: 046104.

21. Abdol MA, Sadeghzadeh S, Jalaly M, Khatibi MM. Constructing a three-dimensional graphene structure via bonding layers by ion beam irradiation. Sci Rep 2019; 9: 8127.

22. Beams R, Cancado LG, Novotny L. Low temperature Raman study of the electron coherence length near graphene edges. Nano Letter 2011; 11(3): 1177-1181.

23. Malard LM, Pimenta MA, Dresselhaus G, Dresselhaus MS. Raman spectroscopy in graphene. Phys Rep 2009; 473(5-6): 51-87.

24. Eckmann A, Felten A, Mishchenko A, Britnell L, Krupke L, Novoselov KS, Casiraghi C. Probing the nature of defects in graphene by Raman spectroscopy. Nano Letter 2012; 12(8): 3925−3930.

25. Malekpour H, Balandin AA. Raman-based technique for measuring thermal conductivity of graphene and related materials. J Raman Spec 2018; 49(1): 106–120.

26. Marton D, Bu H, Boyd KJ, Todorov SS, Al-Bayati AH, Rabalais JW. On the defect structure due to low energy ion bombardment of graphite. Surf Sci 1995; 326(3): L489-93.

27. Singh RI, Mebel AM, Frenklach M. Oxidation of graphene edge six and five member rings by molecular oxygen. J Phys Chem A 2015; 119(28): 7528−7547.

28. Stobinski L, Lesiak B, Malolepszy A, Mazurkiewicz M, Mierzwa B, Zemek J, Jiricek P, Bieloshapk I. Graphene oxide and reduced graphene oxide studied by the XRD, TEM and electron spectroscopy methods. J Elec Spec Rel Phenom 2014; 195: 145-154.

29. Kim SW, Kim HK, Lee K, Roh KC, Han JT, Kim KB, Lee S, Jung MH. Studying the reduction of graphene oxide with magnetic measurements. Carbon 2019; 142: 373-378.

30. Kong L, Bjelkevig C, Gaddam S, Zhou M, Lee YH, Han GH, Jeong HK, Wu N, Zhang Z, Xiao J, Dowben PA, Kelber JA. Graphene/substrate charge transfer characterized by inverse photoelectron spectroscopy. J Phys Chem C 2010; 114(49): 21618-21624.